\documentclass[twocolumn,amsmath,showpacs,amssymb,aps,nofootinbib,floatfix]{revtex4-1}

\usepackage{epsfig}
\usepackage{latexsym}
\usepackage{color}
\usepackage{slashed}
\newcommand{\req}[1]{Eq.\,(\ref{#1})}

\newcommand{\beq}{\begin{equation}}
\newcommand{\eeq}{\end{equation}}
\newcommand{\nn}{  \nonumber \\ }

\newcommand{\GF}{\frac{G_F}{\sqrt{2}}}

\newcommand{\cO}{\mathcal{O}}

\newcommand{\order}[1]{ \mathcal{O} \left( #1 \right) }

\begin{document} 
\hbadness=10000

\title{Neutrino mixing in accelerated proton decays}

\author{Dharam Vir Ahluwalia$^a$, Lance Labun$^b$, Giorgio Torrieri$^c$}
\affiliation{$^a$ Manipal Centre for Natural Sciences,
High Energy Physics Group,
Manipal University,
Karnataka, Manipal 576104, India  \\
$^b$ Department of Physics, University of Texas, Austin, 78712, USA\\
$^c$ IFGW, Universidade Estadual de Campinas, Campinas, S$\tilde{a}$o Paulo, Brazil}

\date{\today}

\begin{abstract}
We discuss the inverse $\beta$-decay of accelerated protons in the context of neutrino flavor superpositions (mixings) in mass Eigenstates.  The process $p\rightarrow n \ell^+ \nu_\ell$ is kinematically allowed because the accelerating field provides the rest energy difference between initial and final states.  The rate of $p\to n$ conversions can be evaluated in either the laboratory frame (where the proton is accelerating) or the co-moving frame (where the proton is at rest and interacts with an effective thermal bath of $\ell$ and $\nu_\ell$ due to the Unruh effect).  By explicit calculation, we show the rates in the two frames disagree when taking into account neutrino mixings, because the weak interaction couples to charge eigenstates whereas gravity couples to neutrino mass eigenstates \cite{grf}.   The contradiction could be resolved experimentally, potentially yielding new information on the origins of neutrino masses.
\end{abstract}


\maketitle

\section{Introduction}
Neutrino mixings are an important piece of evidence for beyond standard model physics.  The effect is phenomenologically described by the fact that neutrinos interact weakly with other standard model particles in flavor eigenstates $|\nu_{\ell}\rangle$ ($\ell=e,\mu,\tau$) that are superpositions of mass eigenstates $|\nu_i\rangle$ determined by a mixing matrix \cite{pontecorvo,neutrev}:
\begin{equation}
\label{mixing}
|\nu_{\ell}\rangle=\sum_i U_{\ell i}|\nu_i\rangle
\end{equation}
with $U_{\ell i}$ known as the  Pontecorvo-Maki-Nakagawa-Sakata (PMNS) matrix.
However, the fact that the mass basis and interaction basis cannot be simultaneously diagonalized means the neutrino weak current and energy momentum tensor do not commute, and hence neutrino number and mass-energy are not commuting observables.  This poses  questions for their coupling to gravity \cite{dharam1,Ahluwalia:1997hc,dharam2,Chryssomalakos:2002bm}.  

To explore the neutrino-gravity coupling, we can perform gedanken experiments involving neutrinos in accelerated frames, using the fact that gravity is represented as invariance under general coordinate transformations, under which $h_{\mu\nu}\to h_{\mu\nu}+\partial_\mu \epsilon_\nu+\partial_\nu\epsilon_\mu$ for arbitrary infinitesimal vectors $\epsilon_\mu$ \cite{weinberg}.  Specifically we can compare conventional Minkowski space quantum field theory calculations to curved-spacetime calculations by noting that transforming the Minkowski metric  by $\epsilon_\mu= a^{-1} (a\tau,0,0,1)$ yields the Rindler metric, representing an observer undergoing constant acceleration $a$.  
One might expect to obtain the same observables since the Minkowski and Rindler frames are related by a symmetry transformation of the theory, given that electroweak theories can be rewritten in a generally covariant way \cite{sewell,biso,vanzella1,matsasrev}.  However, neutrino mixing violates conditions of the construction because, by breaking the mass superselection rule, neutrinos are not representations of the Lorentz group with a well-defined invariant $P^2$.  By considering in detail the results of calculations in different frames, we may be able to gain information or design experiments providing information on the physics of neutrino mixing \cite{grf}.

We consider the inverse $\beta$-decay process $p\to n\nu_e\bar e$,  because it is potentially experimentally observable, but the effect is general and applies to any weak scattering occurring in an accelerated state.  In the ``laboratory'' frame in Minkowski coordinates, the proton is accelerated by an external field and emits the electron and neutrino \cite{vanzella1,vanzella2,suzuki}.  The energy for the process is provided by the accelerating field, and the interaction is the electroweak vertex producing neutrinos in flavor eigenstates.  In the co-accelerating frame in Rindler coordinates, the proton is at rest and interacts with neutrinos and electrons in Rindler states \cite{Fulling,Davies}, which display an effective thermal weight \cite{unruh} and are mass eigenstates, as we will show.  However, mass eigenstates are related to flavor eigenstates by Eq. \ref{mixing}.  Because the neutrino flavor eigenstate is fixed by the electron $\ell=e$, we are forced in this frame to include the PMNS matrix in the amplitude.  The PMNS matrix cannot be factored out, and the rates in the two frames differ.  In this way, basic questions in quantizing neutrino fields \cite{blasone,kopp} are highlighted in the context of quantum field theory in curved spacetime \cite{BandD,matsasrev}. 

Consistency between coordinate frames has been explicitly verified in several cases: the Sokolov-Ternov effect \cite{Sokolov,STexp,Bell:1982qr}, the emitted power in classical electromagnetic radiation \cite{higuchi1,higuchi2} and the rate of $p\to n\nu e^+$ in the absence of neutrino mixing \cite{vanzella1,vanzella2,suzuki}.  Of these, only the Sokolov-Ternov effect has been seen in experiment \cite{STexp}, though it may now be possible to look for the Unruh effect in the electromagnetic radiation of electrons accelerated by high intensity lasers \cite{Chen:1998kp,Schutzhold:2008zza}. Equivalence of observables is expected because the Unruh effect can be seen as a ``Coriolis force'' of quantum field theory in an accelerated state \cite{sudarsky}: an artifact of the coordinate system necessary to restore consistency, its derivation requiring only Lorentz symmetry and quantum mechanics \cite{biso,sewell}.  

Our calculation implements the same technology for treating quantum processes in accelerated frames as preceding work~\cite{vanzella1,vanzella2,suzuki}.  We show that the limit of vanishing neutrino mass or trivial mixing $U_{\ell i}\to\delta_{\ell i}$ recovers these results and the disagreement between frames arises only upon introducing the mixing.  Although the mathematical origin of the disagreement is easily traced to the noncommutativity of weak and energy-momentum currents,  to understand the physics and use the experiment to learn something about neutrino mixing, we study the physics conditions required for the calculation to be valid.

The equivalence of observables is shown to hold in the small acceleration limit, where the response of the system is linear and calculated from the classical state.  We must be able to neglect both classical and quantum backreaction on the accelerating field, such as radiation reaction in the electromagnetic case \cite{lancerad,DiPiazza:2011tq}  (beyond the linear regime the effect of horizons needs to be considered \cite{gteq,boulware,saa}).  This takes the form of a semiclassical approximation, and consequently, the Hawking-Unruh effect has been compared to the Schwinger mechanism of particle production in classical electromagnetic fields \cite{matsasrev,PauchyHwang:2009rz,Labun:2012jf}.  By detailing the approximations required to match the inertial frame result to the accelerated-frame result, we highlight several important differences, in particular how the details of the accelerating field are lost in the Unruh-type calculation.

In the semiclassical limit gravity, described by the classical metric perturbation field $h_{\mu\nu}$, is required by Lorentz invariance to couple only to the conserved energy momentum tensor $T^{\mu\nu}$ \cite{weinberg}.  As usual, we assume the neutrino is well-described as a free particle plus interactions, meaning that in the limit of no weak interactions its energy momentum tensor is that of three free fermion fields in their $\hat{T}_{\mu \nu}$ (ie mass) Eigenstate
\beq
\hat{T}^{\mu\nu}\equiv\frac{\delta L}{\delta h_{\mu\nu}}=\sum_{i=1}^3\left[\bar\nu_iiD^{(\mu}\gamma^{\nu)}\nu_i+g^{\mu\nu}\bar\nu_i(i\slashed{D}-m_i)\nu_i+ \ldots\right]
\label{tdef}\eeq
where $D$ is the covariant derivative, comprising a spinor affine connection defined in \cite{Greiner:1985ce}.
Eq. \ref{tdef}  is diagonal in the mass basis, with mass eigenstates denoted by $i=1,2,3$.    
 At low energies $\ll M_W$, the neutrino interactions are well-described by the Fermi effective theory.  In all, the effective Lagrangian is
\begin{align} 
\label{effneutl}
L=\:& \sum_{i=1}^{3} \bar{\nu}_i (i \gamma_\mu D^\mu -m_i) \nu_i \\ \notag 
& 
+\sum_{\ell}\bar\psi_\ell(i\gamma_\mu D^\mu-m_\ell)\psi_\ell
   +\sum_{\ell=e,\mu,\tau} G_F \hat{J}_{Lh,\mu} \hat{J}_{L\ell}^\mu
\end{align}
where $J_L$ is the left-handed current with nucleons $h=p,n$ for proton and neutron.  For leptons $\ell=e,\mu,\tau$, the weak-charge current is related to the neutrino mass eigenstates by 
\beq
\hat J_{L\ell}^{\mu}=\sum_iU_{\ell i}\: [\bar\psi_\ell\gamma^\mu\nu_i]
\eeq
$G_F=1.16\times 10^{-5}$ GeV$^{-2}$ is the Fermi constant.
Its equation of motion would show the metric field is sourced by the expectation value $\langle\hat{T}^{\mu\nu}\rangle$, but here the kinetic term for the metric is neglected since we take the semiclassical approximation and backreaction is negligible as long as the acceleration remains small.  
The lepton modes are determined including the classical field to all orders by solving the Dirac Hamiltonian, which  is identical to the energy in a local rest frame, $\hat T^{00}$.
The weak current  $J^{\mu}_L$ is conserved, but, because of the PMNS matrix, it does not commute with $\hat{T}_{\mu \nu}$, generating a potential source of tension with gravitational physics even in the semiclassical limit\footnote{In ordinary quantum mechanics, experiments such as Stern-Gerlach show that mutually non-commuting interactions do not give these problems,  because different boundary conditions are generally related by rotations in Hilbert space.  In quantum field theory, as we will see in the two subsequent sections, different boundary conditions give unitarily inequivalent representations. The tension between the gravitational and weak interaction terms, analogous to frustration in condensed matter systems, is therefore apparent only with a full quantum field theoretical treatment.  }.

In the following, we discuss the calculation in each frame, drawing attention to the approximations involved in the low acceleration limit and the necessity of the mass eigenstates and PMNS matrix in the Rindler frame.  We shall perform the calculation in 1+1 spacetime dimensions for notational simplicity; the additional transverse directions do not affect the reasoning for needing the PMNS matrix in the Rindler frame.  Then in Sec.\,\ref{sec:exp} we discuss experimental possibilities for resolving this paradox, and in Sec.\,\ref{sec:concl} we conclude with some broader lessons from this study.

\section{Neutrino mixing in different frames}
\subsection{Inertial frame small $a$ expansion}\label{sec:labframe}

\begin{figure}[t]
\includegraphics[width=120pt]{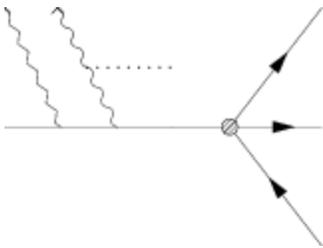}
\caption{\label{diaginertial}Diagram for the process $p \rightarrow n \ell^+\nu_\ell$ in the inertial frame, with the wavy line corresponding to the classical potential, treated to all orders as indicated by the ellipsis. }.
\end{figure}

In the Minkowski frame (Fig.~\ref{diaginertial}), the calculation  amounts to electron and neutrino production by a classical source \cite{Peskin}, because the Fermi theory current-current interaction is treated with a classical hadronic current $\hat J^\mu_{L\ell}\hat J_{Lh,\mu}\to \hat J^\mu_{L\ell}J_{Lh,\mu}^{(cl)}$ \cite{vanzella1}. The calculation is ``semiclassical'' in that quantized lepton fields are produced by a classical source. The time dependence of the source $J_{Lh,\mu}^{(cl)}$ includes an oscillating phase $e^{i(m_n-m_p)t}$ contributing to the energy of the outgoing particles.  To understand better what the process can reveal about the Unruh effect in presence of neutrino flavor mixings, we examine the physics conditions necessary to reduce the hadronic current operator to a classical current, and this begins with an expansion in the small ratios of the acceleration to the other momentum scales $a/m_W,a/m_p,a/\Delta m$, where $\Delta m=m_n-m_p$. 

To consider the proton motion prescribed, the acceleration must be small $a\ll m_p$, so that radiation reaction and backreaction on the external field is negligible.  In fact, the proton's compositeness gives a more stringent constraint since electromagnetic fields $eB,eE\simeq m_\pi^2$ affect its structure, so we require $a\ll m_\pi^2/m_p\sim 19$ MeV.  This implies also $a\ll m_W$ so that the Fermi effective theory is applicable.  The limitation to $a/\Delta m\ll 1$ will become clear later.
The perturbative scale is $a$ and hence the integration volume is $d^4x\sim a^{-4}$, consistent with being the physical scale of the saddlepoint in the tunneling potential for the leptons, which is localized (at the scale $a^{-1}$) to the region around the turning point.\footnote{In any case, at longer wavelengths the proton motion is perturbed by the emission of radiation, which according to classical formulae has a wavelength of order $a^{-1}$.}  For the same reason, the acceleration should be constant over $\sim a^{-1}$ in order for the semiclassical approximation to apply for the leptons, which have be Broglie wavelengths of order $\Delta m$.


The $p\to n\nu e^+$ operator from the lagrangian interaction in \req{effneutl}
\begin{equation}\label{bareM}
\cO=4\GF(\bar\Psi_n\gamma^\mu P_L \Psi_p)(\bar\nu_e\gamma_\mu P_Le),\quad P_L=\frac{1-\gamma_5}{2}
\end{equation}
is a perturbative interaction, and the Fermi theory suffices since the momenta involved are all $\ll m_W$.
To match the Rindler frame calculation which involves a correlation function restricted to a classical trajectory, we must reduce the proton to a point-like classical particle.  This differs from the semiclassical approximation in field theory, which entails solving the Dirac or Klein-Gordon equation in the presence of the classical accelerating  potential \cite{schwinger,dunne}.  To reduce to a classical trajectory, the anti-particle components of the proton wavefunction are integrated out first by using translational invariance of the quasi-constant external field to focus on the region around the turning point in the trajectory where the proton is nonrelativistic. 
Since $a\ll m_\pi$, the hadron is nonrelativistic for a time of order $a^{-1}\ln m_n/m_\pi>a^{-1}$, found by solving for the time for the proton to gain $1m_p$ of energy.  In this frame, the proton and neutron are heavy fermions, allowing us to extract the mass as the large part of the phase and integrate out the antiparticle components of the spinor
\begin{equation}
\Psi_p\to\sum_v h_p(v)e^{-im_pt}, \Psi_n^*\to\sum_{v'} h_n(v')e^{im_nt}
\end{equation}
Corrections to the dynamics of the $h_v$ fields as well as antiparticle components are suppressed by $1/m_p$.
Also at leading order in $1/m_p$, we can neglect the recoil, which means we drop the residual nucleon momenta and set $v=v'$.  Considering the ``coherence time'' of the process to be $\delta t\sim 1/m_W$, the velocity change during the interaction is $\delta v=a\delta t\sim a/m_W$, also a subleading correction.
 
As the nucleons are nonrelativistic particles with negligible recoil, the matrix element of the current satisfies the classical equation of motion in the external potential and therefore matches to the classical trajectory $\xi_\mu(\tau)$ (essentially the Ehrenfest theorem)
\begin{equation}
e^{-i\Delta m t}\langle n|h^*_n(v)\gamma^\mu P_L h_p(v)|p\rangle \to J^\mu_{h,cl}(x)
\end{equation}
with 
\begin{equation}\label{Jhcl}
J^\mu_{h,cl}(x)=\frac{1}{2}e^{-i\Delta M t}\frac{u^\mu(x)}{u^0(x)}\delta(x_i-\xi_i(\tau))
\end{equation}
The charge is included in the coupling constant $G_F$ factored out of the fermionic currents.  Here, the factor $1/2$ comes from projecting an unpolarized current onto lefthanded states.   

Evaluation of the matrix element proceeds straightforwardly, and the differential rate is
\begin{align}
\label{dwd4k}
\frac{dW_{\rm Mink}}{d^2k d^2k'}=&\: G_F^2\!\!\int\! d\tau d\tau'e^{i\Delta M (t-t') + i(\omega_e+\omega_\nu')(t-t')-i(\vec k+\vec k')\cdot(\vec\xi-\vec\xi')} \nn
 & \times e^{}(k_\mu k_\nu'+k'_\mu k_\nu-g_{\mu\nu}k\cdot k')u^\mu(\tau)u^\nu(\tau')  \nn
 &+\mathrm{terms~odd~in~}k,k'
\end{align}
We have used $dt/u_0(\tau)=d\tau$.  We will discuss the phase space below, which comprises dependence on the final state lepton masses.  Terms odd in $k,k'$ vanish under the phase space integral and will be dropped in the following.

The accelerated trajectory is parameterized as
\begin{align}
\xi_\mu(\tau)=a^{-1}(\sinh a\tau,0,0,\cosh a\tau), \quad u_\mu=\frac{d\xi_\mu}{d\tau}
\end{align}
The reason for the finite result for a process seemingly violating energy conservation is in this noninertial trajectory: 
  The process is kinematically allowed due to the acceleration of the proton, because the external potential provides the rest energy difference between initial and final states.  As long as the potential is ``weak'' $a\ll \Delta m$, the process is exponentially suppressed, just as is spontaneous pair production in quasi-constant electric fields \cite{schwinger,dunne}.  

The phase factor depends on the coordinates $x^\mu$ with spatial coordinates replaced by $\xi^i(\tau)$ of the trajectory.  Expanding around the origin where the hadron is instantaneously at rest, $\tau=t-\Delta x\simeq t(1-at/8)$ to leading order in $at$. Since the phase factor enforces energy-momentum conservation, $t\sim \Delta m^{-1}$ and we have $t-t'=\tau-\tau'$ to leading (zeroth) order in $a/\Delta m$.  

Making the substitution $t-t'\to \tau-\tau'$, the integrand has no dependence on $\bar\tau=\tau+\tau'$, and one can define a rate per unit (proper) time.  Defining $T=\int d\bar\tau$ we have
\begin{align}
\frac{1}{T}\frac{dW_{\rm Mink}}{d^2k d^2k'}=& \frac{1}{T}\frac{dW}{d^2\tilde{k} d^2\tilde{k}'} \nn 
=&\:\:G_F^2\!\!\int_{-\infty}^{\infty}\!\!\!d\sigma\: e^{i 2a^{-1}(\tilde k_0+\tilde k_0')\sinh \sigma a } \nn
 &\times\big(\tilde k_0\tilde k_0'+\tilde k_z\tilde k_z'\big)
\label{tausubbedrate}
\end{align}
where the rescaled variables are 
\begin{align*}\notag
\tilde k_0=\:&\: k^0\cosh \bar\tau a- k_z\sinh \bar\tau a\\
\tilde k_z=\:&\: k^z\cosh \bar\tau a- k_0\sinh \bar\tau a
\end{align*}
We note that $\tilde{k}$ and $k$ are related by a Lorentz transformation and, provided the phase space is Lorentz invariant, the rate $W_{\rm Mink}$ in Eq. \ref{tausubbedrate} is also Lorentz invariant, defined in terms  of the conversion reaction in its proper time.   The transformation properties of the rate are explicit in $T^{-1}$ and thus $W_{\rm Mink}/T$ is directly comparable to the rate obtained in the comoving frame.  The rate measured in the lab frame is obtained by convolving with the proton Lorentz factor along its lab-frame trajectory \cite{vanzella1} as in the consideration of decays high energy particles.

We discuss the phase space integration and comparison to the Rindler frame rate after obtaining the Rindler rate.


\subsection{ Comoving frame calculation \label{sec:comoving}}

The coordinates in the comoving frame are given by the Rindler metric, which in the right wedge $z>|t|$ has the form $ds^2=\tilde g_{\mu\nu}d\tilde x^\mu d\tilde x^\nu=u^2dv^2-du^2$.  The spacelike coordinate  $u$ is related to the Minkowski coordinates by $u^2=z^2-t^2$ ($0<u<\infty$) and determines the magnitude of the acceleration of the trajectory relative to the Minkowski space.  We assign a continuous momentum eigenvalue $\omega$ conjugate to the time-like coordinate $v$, and the corresponding wavefunctions are determined by solving the Dirac Hamiltonian \cite{Greiner:1985ce}
\begin{align}\label{RindlerDiraceqn}
\omega\psi\equiv i\partial_v\psi=\hat H\psi=
\gamma^0(\gamma^3u^{1/2}\partial_u u^{1/2}-m_\ell)\psi
\end{align}
for each lepton $\ell=e,\nu_i$.  The $\tilde\gamma$ are the Rindler frame Dirac matrices, satisfying $\{\tilde\gamma^\mu,\tilde\gamma^\nu\}=2\tilde g^{\mu\nu}$.  Note that the Hamiltonian operator is identically the $\mu=\nu=0$ element of the energy-momentum tensor operator.

The Rindler space field operators are thus expanded
\begin{align}
\chi(x)=\int \frac{d\omega}{2\pi}\sum_\sigma\left(e^{-i\omega v}\hat b_{\sigma\omega}\chi_{\omega,\sigma}(u)+e^{i\omega v}\hat d^\dag_{\sigma\omega}\chi_{-\omega,-\sigma}(u)\right)
\end{align}
in terms of the eigenmodes 
\begin{align}\label{rindlermodes}
&\chi_{\omega,+}=N^{1/2}
\left(\begin{array}{c}
i\Phi^-_\omega+\Phi^+_\omega \\
0 \\
i\Phi^-_\omega-\Phi^+_\omega \\
0
\end{array}\right), \quad
\chi_{\omega,-}=i\gamma^0\gamma^2\chi_{\omega,+} \nn 
&\Phi^{\pm}_\omega=K_{i\frac{\omega}{a}\pm\frac{1}{2}}(m_\ell u), \quad
 N_{\omega}^{-1}=\frac{2m_\ell}{a}\cosh\frac{\pi\omega}{a}
\end{align}
The Rindler frequency takes all real values with no mass gap, and the mass appears in the wavefunction.  It is the intrinsic dependence of the wavefunction on the mass that makes the sum over mass eigenstates \emph{not} factorizable.
For each field, the electron and each neutrino mass eigenstate, a corresponding complete set of solutions covers the right spacelike wedge.

\begin{figure}[bt]
\includegraphics[width=80pt]{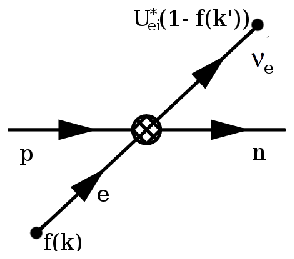}
\includegraphics[width=80pt]{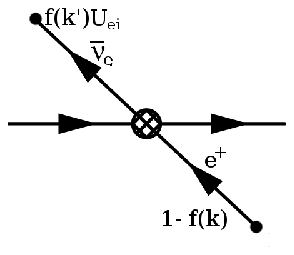}
\includegraphics[width=80pt]{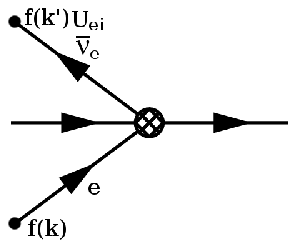}
\caption{\label{diagcomoving} Diagram for the process $p \rightarrow n \ell^+\nu_\ell$ in the comoving frame,  where the electron and neutrino legs are connected to thermal distributions. }   
\end{figure}

Forming the matrix elements for the processes \req{comovingprocs}, we take the Minkowski in and out states, bringing in the Bogoliubov coefficients relative to the Minkowski particle $\psi^+$ and antiparticle $\psi^-$ modes
\begin{align}
&\chi_\omega=\alpha_\omega\psi^++\beta_\omega\psi^-\\
\alpha_\omega=&\frac{e^{\pi\omega/2a}}{(2\cosh(\pi\omega/a))^{1/2}},~~~\beta_\omega=e^{-\pi\omega/a}\alpha_\omega
\end{align}
This shows that  in the Rindler frame, the $p\to n$ transition is matched to three distinguishable processes
\begin{align}
(I)\qquad &p\,e \underset{T_{U}}{\longrightarrow} n\,\nu_e\, ,
\nn \label{comovingprocs}
(II)\qquad &p\,\bar \nu_e \underset{T_{U}}{\longrightarrow} n\,\bar e \,, 
\\ \notag
(III)\qquad &p\,e \bar \nu_e \underset{T_{U}}{\longrightarrow} n  
\end{align}
corresponding to the absorption by a static proton of, respectively, an electron, a neutrino and both an electron and a neutrino from the Unruh thermal bath (Fig. \ref{diagcomoving}). 

Additionally, the wavefunctions \req{rindlermodes} do not cover all of Minkowski space, even when combined with the corresponding wavefunctions in the left, future and past wedge.  A complete set of wavefunctions, in terms of which the Minkowski modes can be expanded, requires defining $\delta$-function sources on the lightcones $t\pm z=0$ \cite{Soffel:1980kx}.  For neutrinos, these sources are clearly in the mass basis and show that the wavefunctions in the calculations are required to be on-mass-shell solutions to \req{RindlerDiraceqn} at the boundary of the Rindler wedge.  For this reason, we take the observed, asymptotic states in Rindler space to be mass eigenstates, and this requires introducing a factor of $U_{ie}$ to rotate to the electron field appearing in the operator.  Therefore, the squared matrix elements are
\begin{subequations}\label{comovingMs}
\begin{align}
&|i\mathcal{M}_I|^2=\mathcal{J}^{\mu\nu}_{h}(x,x')\sum_i|U_{ei}|^2\mathcal{L}_{\mu\nu}^{(i)}(x,x')|\alpha_\omega^{(e)}|^2|\beta_\omega^{(\nu_i)}|^2 \\
&|i\mathcal{M}_{II}|^2=\mathcal{J}^{\mu\nu}_{h}(x,x')\sum_i  |U_{ei}|^2\mathcal{L}_{\mu\nu}^{(i)}(x,x') |\beta_\omega^{(e)}|^2|\alpha_\omega^{(\nu_i)}|^2\\
&|i\mathcal{M}_{III}|^2=\mathcal{J}^{\mu\nu}_{h}(x,x')\sum_i  |U_{ei}|^2\mathcal{L}_{\mu\nu}^{(i)}(x,x')|\beta_\omega^{(e)}|^2|\beta_\omega^{(\nu_i)}|^2
\end{align}
\end{subequations}
where the hadronic and leptonic tensors are
\begin{align}
\mathcal{J}^{\mu\nu}_{h}(x,x')&=\delta^\mu_0\delta^\nu_0 e^{-i\Delta M (v-v')}\delta(u-a^{-1})\delta(u'-a^{-1}) \\
\mathcal{L}_{\mu\nu}^{(i)}(x,x')&=\sum_{\sigma,\sigma'}[\bar\chi^{(\nu_i)}\tilde\gamma_\mu\tilde P_L\chi^{(e)}](x)[\bar\chi^{(\nu_i)}\tilde\gamma_\mu\tilde P_L\chi^{(e)}]^\dag(x')
\end{align}
in Rindler coordinates.  Note that $\tilde\gamma^5=i\tilde \epsilon_{\mu\nu\kappa\lambda}\tilde\gamma^\mu\tilde\gamma^\nu\tilde\gamma^\kappa\tilde\gamma^\lambda=\gamma^5$.

We note that if  the $U_{ej}^* U_{ei}$ interference terms between mass Eigenstates were also added, inertial and comoving calculations would match, as this would be equivalent to assuming that a charge Eigenstate is the asymptotic state also in the accelerating frame.  
  This would be consistent with the standard inertial-frame model of neutrino propagation, in which the different mass-states co-propagate.  However including the interference terms would violate the Kubo-Martin-Schwinger (KMS) definition of a thermal state of a quantum system \cite{kms,matsasrev} by adding coherent, off-diagonal correlations in the density matrix.  Consequently, the accelerated neutrino vacuum state would not be thermal, contradicting the essential characteristic of the Unruh effect and its formal derivation within quantum field theory \cite{biso,sewell}.
It is possible that nature realizes this case, seeing that exact satisfaction of the KMS definition has not been proven for general interacting theories \cite{matsasrev,truran} but such an alternative would have much more general implications than this work, given that the essence of the Unruh effect is the thermality of the state.

Excluding the  $U_{ej}^* U_{ei}$ interference terms, we have followed the current best prescription for neutrino-emission and accelerated-frame calculations: weak interactions are perturbative, satisfying the condition to construct the usual wavefunctions for neutrinos in the frame of an accelerated observer (the proton) as free-particle solutions to the Dirac equation \req{RindlerDiraceqn}.  As the wavefunctions \req{rindlermodes} have non-perturbative dependence on $m_\nu$, boundary conditions require the asymptotic states in the accelerated frame are on mass-shell and contradict the inertial frame model of neutrino propagation, which does not require asymptotic neutrino states to be on mass-shell and is an experimentally-verified effective description.  The calculations in the inertial and accelerated frames are independently clearly defined, but, as we see now, imply incompatible results.  This incompatibility can be seen as the price of maintaining the KMS condition for a particle in a mass superposition.


Summing the processes $I,II,III$ and integrating, the result has an analytic form similar to \cite{vanzella1}, with the additional sum over neutrino mass eigenstates,
\begin{align}
  W_{\rm Rind}=\frac{G_F^2a}{8\pi^2}e^{-\frac{\pi\Delta m}{a}}&\sum_i \left| U_{ei} \right|^2 \int_{-\infty}^{\infty}d\omega H(\frac{\omega}{a},\frac{m_e}{a},\frac{m_{\nu_i}}{a})
  \label{comovrate} 
\end{align}
where the integrand is
\begin{align}
\label{hdef}
H(\frac{\omega}{a},\frac{m_e}{a},\frac{m_{\nu_i}}{a})=\:&\:
 6|K_{i\frac{\omega}{a}+\frac{1}{2}}(\frac{m_e}{a})|^2|K_{i\frac{\omega'}{a}+\frac{1}{2}}(\frac{m_{\nu_i}}{a})|^2
\\ \notag &
+\mathrm{Re}\left[K_{i\frac{\omega}{a}+\frac{1}{2}}(\frac{m_e}{a})^2 K_{i\frac{\omega'}{a}+\frac{1}{2}}(\frac{m_{\nu_i}}{a})^2\right]
\end{align}
with $\omega'=\omega-\Delta m$ and $K_\nu(z)$ are the modified Bessel functions of the second kind.
The $m_{\nu_i}$ dependence of the integrand prevents factoring out the PMNS matrix and using $\sum_iU_{ei}^* U_{ei}=1$.  Without neutrino mixing, this result reduces to Eq.\,(9) of \cite{vanzella2}.  
 
\subsection{Total rate comparison \label{secphase}}

The simplest observable is the total number of electron-neutrino pairs detected, so that the Minkowski differential rate \req{tausubbedrate} must be integrated over the total interaction time the electron and neutrino momenta in the final state.  As usual, we ensure the 4-momentum integrals converge by putting the electron and neutrino on mass shell.  Since the energy of the final state neutrino is $E_\nu\sim\Delta m\gg m_\nu$, it is a good approximation to the neutrino phase space to treat it as massless,
\begin{equation}
\label{totrate}
W_{\rm Mink}=\int d^2k d^2k'\frac{dW}{d^2kd^2k'} \delta(k^2-m_e^2)\delta\!\left((k^\prime)^2\right)
\end{equation}
The result converges to the $m_\nu\to 0$ limit given in \cite{suzuki}
\begin{equation}
\label{inertrate}
W_{\rm Mink}= \frac{G_F^2a}{8\pi^2}e^{-\frac{\pi\Delta m}{a}}
\int d\omega H(\frac{\omega}{a},\frac{m_e}{a},0)
\end{equation}
with $H(x,y,z)$ the same function defined in \req{hdef}.  Obtaining this form requires the integral transformations in \cite{suzuki} and the identity for the Meijer G-function
\begin{widetext}
\begin{equation}
x^\sigma K_\nu(x)K_\mu(x)= \frac{\sqrt{\pi}}{2}
G^{40}_{24}\left( x^2 \left|
\begin{array}{c}
      \frac{1}{2}\sigma,{\frac{1}{2}}\sigma+{\frac{1}{2}} \\\
        \frac{1}{2}(\nu+\mu+\sigma),
        \frac{1}{2}(\nu-\mu+\sigma),
        \frac{1}{2}(-\nu+\mu+\sigma),
        \frac{1}{2}(-\nu-\mu+\sigma)
\end{array}
\right.\right),
\label{kk}
\end{equation}
\end{widetext}
These results carry through also for non-vanishing neutrino mass, and the results of \cite{vanzella1,suzuki} are reproduced.

The inertial frame rate $W_{\rm Mink}$ Eq. \eqref{inertrate} differs from the comoving rate $W_{\rm Rind}$ \req{comovrate} since $m_\nu$ enters the wavefunctions in the comoving frame.  

The corrections to the integrand in \req{comovrate} due to the $m_\nu$ dependent terms appear to be small, but their existence suffices to preclude factoring out $\sum_iU_{ei}^* U_{ei}$.  
Since elementary particles live in different irreducible representations no matter how small their masses, the identification of the created particle as a superposition of mass eigenstates survives to the infrared limit $V^{1/3}\to\infty$ where the phase space is defined.   
The wavefunctions come with a factor $V^{-1/2}$ (where $V$ is the accessible configuration space volume) and the limit $V \rightarrow \infty$ must be taken with $m_\nu$ finite for the approximation made in Eq. \ref{totrate} to be valid.   Taking $m_\nu\to 0$ first would remove the distinction between flavor and mass eigenstates and there would be no mixing, in contradiction to observations but sufficient to make the inertial and comoving frame rates coincide.  Therefore the massless limit does not commute with the infrared limit, and  setting $m_\nu=0$ in the Rindler frame Eq. \ref{comovrate} would distort the infrared physics that identifies the asymptotic state as a mass eigenstate. 
This noncommutativity of limits is common to situations where infrared physics plays a crucial role (chiral symmetry breaking being an example \cite{smilga}).   Its exact mathematical treatment most likely necessitates consideration of the issues described in the discussion section.

On the other hand, one may worry that the neutrino mass should have been included in the phase space \req{totrate}, and doing so would lead to the inclusion of PMNS factors for the conversion to neutrino mass eigenstates.
However, we recall that a neutrino is created in a flavor eigenstate, mixes over macroscopic distances and is detected in an experiment which again projects a flavor eigenstate. Since the neutrino energy is $\sim\Delta m$, presently conceivable experiments would not be able to resolve the impact of the neutrino mass on the detection scattering event.  This differs from particles in collider experiments, where detection occurs when the outgoing particle interacts with a macroscopic quasi-classical detector\,\cite{colemanhepp}: the quantum evolution is effectively projected to a near-pure momentum eigenstate and reduces to a classical phase space integral in this limit\:\cite{weinberg}.  

As we pointed out in the introduction, this discrepancy is expected because of  the ambiguity inherent in combining non-mass Eigenstate fields with non inertial frames.  In short, taking the $m_\nu\to 0$ limit in the phase space in the Minkowski frame is a kinematical approximation separating corrections of order $m_\nu/E_\nu\ll 1$, whereas the $m_\nu\to 0$ limit in the Rindler frame calculation makes an $\mathcal{O}(1)$ change in the infrared physics by removing mixing.  Deforming to a nonflat metric deforms the mass eigenstate creation and annihilation operators.   In the infrared limit, such operators define on-shell states, yet these are not selected by inertial detectors.   Thus, one cannot expect descriptions in different  frames to give the same scalar observables, as is generally true for interacting field theories.
The resulting difference in the neutrino flavor distribution could in principle be observed by a flavor-sensitive neutrino detector $\sim 1$ oscillation length away from the accelerated-proton source, but we discuss potentially more achievable schemes in the section.

\section{ Experimental opportunities\label{sec:exp}}

This ambiguity warrants experimental investigation.  
  A significant advantage to considering the inverse-beta-decay process is that it is reasonable to ask whether and how it could be experimentally verified.
To test the above effect, we need only to accelerate protons and observe the rate of correlated neutron-positron emission.  Although the physics is contained in neutrino fields, no neutrinos need to be detected.  In contrast, weak processes with lower threshold, such as neutrino bremsstrahlung from an accelerated electron, require detecting the neutrino(s) in the final state.  The authors know of no current plans for facilities where such an experiment is possible.  Accelerated proton decay may be achievable with next generation high intensity lasers.

The challenges in observing this effect are (1) achieving sufficiently high $a$ to overcome exponential suppression, and (2) maintaining that acceleration for sufficiently long that the quasi-constant approximation applies.  The first is a matter of having a sufficiently strong and uniform field, something which is currently being developed in many contexts within pure and applied physics.  The second may be partially offset by studying more general non-constant accelerated trajectories; however, in all cases the duration of the acceleration must be much larger than the ``equilibration time'' to the noninertial vacuum \cite{Doukas:2013noa}.  As a conservative estimate $\Delta t\lesssim (\alpha m_\nu)^{-1}\sim 6\,\mu$m=2 fs, though there is significant dependence on the acceleration profile.  The field must also do work on the accelerated particles seeing that the final state rest energy is $\simeq \Delta M$ greater than the initial state. 

Only electromagnetic fields can provide high accelerations for durations comparable to $\Delta t$.  In the lab, these are provided by high intensity lasers.  The current generation of facilities, such as the Texas Petawatt \cite{petawatt} and the Extreme Light Infrastructure (ELI) \cite{dunne,eli} create $L\sim20-40\lambda\simeq 20-40\mu$m pulses and achieve fields of $eE=(10^{-4}-10^{-3})m_e^2$.  A proton entering a field of this strength is accelerated by $a\simeq 10^{-6}$\,MeV, which means the rate $\propto e^{-\pi \Delta m/a}$ is effectively zero.  Next generation facilities are under discussion as part of a longer program to bring the QED critical field strength $eE_c = m_e^2$  within reach of laboratory experiment.  In a field of QED-critical strength, a proton's acceleration is $a\sim 10^{-3}$\, MeV, still a large suppression.  If this process is to be experimentally realized, additional analytic study would be necessary, including likely nonperturbative sub-threshold enhancements, as have recently received much study in the context of spontaneous electron-positron pair creation, see for example \cite{enhancement} and related references.

By comparison, other long wavelength fields are: (1) the mean electrostatic field inside an LHC bunch, estimated  $eE\sim (\mathrm{keV})^2$, using a cylinder with dimensions of the bunch ($N=10^{11}$, bunch radius $10^{-5}$ cm and length $L\sim 10$\,cm); (2) mean fields inside fixed targets at hadron colliders $eE\sim (100 \mathrm{eV})^2$ again with $L\sim 10$\,cm; and (3) plasma wakefield and crystal accelerator concepts $eE\lesssim 10^{-6}m_e^2$ over $L\sim\mu$m.  Larger $L$ in the LHC bunch is not sufficient to overcome the relative $e^{-10^3}$ suppression.

Quantum effects, in particular electron-positron pair production would be a large background.  Although pair production is similarly exponentially suppressed, the lower threshold $2m_e<\Delta m$ means it would occur at higher rate.  For this reason, it would be necessary to solve the positron dynamics in the field to be able to correlate them with emitted neutrons.  
 Scattering with electrons present in the system would also be a background, to be eliminated with $ne^+$ correlations and the known $p e \rightarrow n \nu$ scattering cross-section.  The theory work to understand these particle dynamics and backgrounds to greater precision is under way as part of the program to observe strong field QED effects in forthcoming laser experiments.

\section{Discussion\label{sec:concl}}

We close with a discussion of what we expect to see if such an experiment is ever performed.
The ambiguity between the inertial frame calculation and the comoving calculation \cite{vanzella1}, due to an interplay of mass and mixing ($U_{\ell j}$) terms in Eqs \ref{totrate} and \ref{comovrate}, cannot be removed by a coordinate transformation.  
Either the comoving or the inertial calculation or both will give a wrong result, and we elaborate a few of the options revealed by making explicit the assumptions in the calculations.

First, by writing the lagrangian \req{effneutl}, we consider the neutrino mass a tree-level operator, unmodified by the interaction producing the acceleration.  This is consistent with the neutrino mass being either a ``fundamental'' operator or generated by other interactions that have been integrated out, such as coupling to the Higgs or high scale beyond Standard Model physics.  Measuring rates in agreement with the inertial frame calculation would provide evidence that the neutrino mass is an effective operator of one of these types.  Specifically, this treatment in the inertial frame is appropriate as long as the cutoff scale for the (effective) mass operator is $\gg a$. Matching the comoving frame calculation to the observed rate in this case requires deeper understanding how the Higgs condensate transforms under the noninertial coordinate change or how effective operators with finite cutoff scales are represented in the curved space.

It is also possible that neither calculation is an accurate prediction.  Clearly, the model here is a crude approximation to realistic experimental conditions using electromagnetic fields to accelerate the proton: neutrons have zero charge and could not continue on the prescribed accelerated trajectory as the proton and the electron should be treated nonperturbatively in such a strong electromagnetic field.  In particular, conventional low energy (electromagnetic) initial and final state radiation can significantly modify the rate, needing to be resummed as is expected for high energy processes in high intensity lasers.

An essential approximation is that the source of particles is classical, though the radiated particles are treated as quantum, and we argued this approximation is valid when the acceleration is the smaller than any of the energy scales associated with that current or the interaction.  
In this respect, the Unruh effect and specifically accelerated $p\to n\nu_e e^+$ conversion are semiclassical calculations and similar to spontaneous pair production in classical electromagnetic fields.  However, there is a crucial difference between the Unruh effect and the Schwinger effect: in the Unruh effect, the acceleration has been reduced to a classical accelerated trajectory, and the details of the accelerating field have been lost. All that one requires is a classical trajectory whose quantum fluctuations are minimized, so the action is real and close to the extremum.  The Schwinger effect, on the other hand, necessitates an $\mathcal{O}(\hbar)$ correction to the action. 

This removal of the information about the acceleration is consistent with the original calculation in \cite{unruh} which requires an essentially classical detector capable of projecting single particle eigenstates.  The nucleon here can be treated as a classical detector because distinguishing a neutron from a proton is ``easy'' for a classical macroscopic detector, e.g. a particle tracker in a magnetic field.

That said, both the Schwinger effect and the Unruh effect rely on the approximation that quantum fluctuations, controlled by $\hbar$, are small.  In the Schwinger effect, the occupancy number of the classical electromagnetic field is large $\gg 1$ implying that the field's action is $\gg\hbar$, while in the Unruh effect it means $a/\delta p \gg \hbar$ where $\delta p$ is the typical microscopic momentum exchange producing the acceleration.  When the acceleration is produced by a field these two conditions are equivalent.  On the other hand, too large $a$ violates the assumption of no backreaction.

The ambiguity highlighted here derives from breaking the mass superselection rule in the Lorentz group.  Since the gravitational field must be sourced by the energy-momentum tensor,
invariance of the $h_{\mu\nu}T^{\mu\nu}$ coupling requires the particles states in the accelerated frame are mass eigenstates, which is the content of solving the Dirac equation in the Rindler coordinates \req{RindlerDiraceqn}.  In the inertial frame, acceleration provides the energy required for the reaction but the interaction term always projects out flavor eigenstates.   

This ambiguity does not necessarily contradict the fact that the Unruh effect arises out of axiomatic field theory when written with general covariance \cite{biso,sewell}.  If it were possible to incorporate neutrino mixing into standard quantum field theory with neutrino fields as well-defined representations of the Poincar\'e group, then there would be a contradiction.  However, the models of neutrino mass and mixing in the literature all invoke interactions, for example with Higgs-like condensates or with high-scale particles integrated out.  

No axiomatic construction such as found in \cite{biso,sewell} is known for interacting theories, which are defined only perturbatively, so care needs to be taken in applying conclusions from axiomatic field theories to theories with non-trivial vacua.  It is well known that condensates break general covariance to zeroth order: in gravity, a Higgs-like mechanism is a common method to introduce consistently additional gravitational degrees of freedom \cite{massivegrav}, or see \cite{pappalardo} for an example in the context of higher-dimensional extensions of the Standard Model.  The full symmetry is restored only by constraining the higher-order interactions among new degrees of freedom on these backgrounds.  Generally, low-energy effective actions exhibiting neutrino mixing will not satisfy these constraints, and therefore we must determine the transformation of the condensates and interactions under general coordinate changes before we can say how the effective interaction appears in the Rindler frame. 

Hence, the result presented here motivates systematic study of how effective interactions with high-scale particles or condensates transform, or what conditions are necessary for the interactions to be consistent with general covariance.
If neutrino masses originate in high-scale physics giving rise to low-energy effective interactions,
the inertial frame tree-level calculation would be protected by the fact that the scale $\Lambda$ responsible for the mixing (the Higgs condensate in the Weyl case, the non-renormalizeable operator in the seesaw case) is much larger than any momentum scale associated with the classical field.  The effective theory in which the inertial frame calculation was performed should therefore be correct to $\order{a/\Lambda \ll 1}$.   In the comoving frame, however, the tree level calculation would be based on an incorrect approximation since the neutrino would not be a point-like particle but rather a composite between a ``bare'' degree of freedom and a zero-momentum condensate. 
In this case, the experimentally measured conversion would follow the inertial calculation, and an experimental realization of this paradox would teach us about the origin of the neutrino mass.


Another simple solution to the paradox is that the mixing matrix vanishes in the UV,   just as for Kaons\footnote{Kaons are mass-degenerate Eigenstates at the precision level the mass of an unstable particle can be conceivably measured \cite{christ}. They oscillate because flavor does not commute with the weak isospin charge.  Neutrinos oscillate because weak isospin charge does not commute with mass, as this work discusses.   In the Kaon case, the mixing is due to an IR operator coupling the Higgs condensate to Fermions, and it is possible a similar mechanism applies in the neutrino case.}. In such a case, neutrinos with energy $\sim\Delta m$ could be mass degenerate.  For consistency, also in the Rindler frame the neutrinos should lose the information of the mass eigenstate, and neutrino states with frequency$\sim\Delta m$ appear as superpositions of the mass states.  No PMNS matrix would be necessary, removing the contradiction.

The proposed experimental program explores a physics domain orthogonal to that usually considered for beyond Standard Model physics:  instead of one scattering event with large momentum transfers we consider a uniform acceleration, which means a large number of interactions with soft quanta.
Since the effective potential giving rise to most, if not all mass terms is thought to be the Higgs mechanism involving a condensate of zero momentum quanta, it is reasonable to suppose new physics will show up in this regime provided the scale of the relevant term of the effective lagrangian is comparable to the acceleration, as it certainly is here.

As a result of this orthogonality, the dependence on neutrino masses and mixing parameters of the rates calculated here can be very different from the usual ones.   If the comoving frame calculation, Eq. \ref{comovrate} applies, it would open up an invaluable laboratory to study neutrino masses and mixing angles directly:  Unlike any other mechanism within known physics, the Rindler boundary conditions ``prepare'' the neutrino in a pure mass Eigenstate.
The reaction rate depends on the absolute value of the phases within the mixing matrix and the absolute value of the masses, rather than mass differences.   The experimental opportunities inherent in this realization are numerous enough that even enumerating them is beyond the scope of this work (For one, performing these experiments with both protons and antiprotons would give the CP-violating terms of the PMNS matrix).  If the inertial frame rate Eq. \ref{inertrate} applies, the prospects of using the processes described here to study the PMNS matrix appear bleaker, since this matrix does not appear in Eq. \ref{inertrate}.   However, precision studies of such a process might be able to shed light of the origin of the neutrino mass term, since loop corrections will certainly be sensitive to UV structure of the operator appearing as a mass in the IR.   A detailed analysis of this is beyond the scope of this paper.

In conclusion, we have argued that the conversion of accelerated protons into neutrons is a promising laboratory to study the origin of neutrino masses and mixings.   If the inertial frame calculation is correct, we do not have sensitivity to the neutrino mass absolute values, unless our choice of the phase space is incorrect.
Admittedly, technologically such experiments are still somewhat out of reach, but we hope that the ingenuity of the QED and intense laser community will make such experiments feasible in our lifetimes.

GT acknowledges support from FAPESP proc. 2014/13120-7 and CNPQ bolsa de
produtividade 301996/2014-8. L.L. was supported by NNSA cooperative agreement DE-NA0002008, the Defense Advanced Research Projects Agency's PULSE program (12-63-PULSE-FP014), the Air Force Office of Scientific Research (FA9550-14-1-0045) and the National Institute of Health SBIR 1 LPT$\_$001.

\end{document}